# MONARC SIMULATION FRAMEWORK

Ciprian Dobre* and Corina Stratan**

* Department of Computer Science, Politehnica University of Bucharest, Automatic Controls and Computer Faculty, Romania, e-mail: cipsm@cs.pub.ro

** Department of Computer Science, Politehnica University of Bucharest, Automatic Controls and Computer Faculty, Romania, e-mail: corina@cs.pub.ro

*Abstract* - *This paper discusses the latest generation of the MONARC (MOdels of Networked Analysis at Regional Centers) simulation framework, as a design and modelling tool for large scale distributed systems applied to HEP experiments. A process-oriented approach for discrete event simulation is well-suited for describing concurrent running programs, as well as the stochastic arrival patterns that characterize how such systems are used. The simulation engine is based on Threaded Objects (or Active Objects), which offer great flexibility in simulating the complex behavior of distributed data processing programs. The engine provides an appropriate scheduling mechanism for the Active objects with support for interrupts. This approach offers a natural way of describing complex running programs that are data dependent and which concurrently compete for shared resources as well as large numbers of concurrent data transfers on shared resources.*

*The framework provides a complete set of basic components (processing nodes, data servers, network components) together with dynamically loadable decision units (scheduling or data replication modules) for easily building complex Computing Model simulations. Examples of simulating complex data processing systems are presented, and the way the framework is used to compare different decision making algorithms or to optimize the overall Grid architecture and/or the policies that govern the Grid's use.*

*Keywords - simulation, distributed systems, architecture validation, design models*

## I. INTRODUCTION

The future Large Hadron Collider (LHC) experiments have envisaged computing models involving many hundreds of physicists doing analysis jobs at institutions around the world. These models encompass a complex set of wide-area, regional and local-area networks, a heterogeneous set of compute- and data-servers, and a yet-to-be determined set of priorities for group-oriented and individuals' demands for remote data and compute resources. Each of the experiments foresees storing and partially distributing data volumes of Petabytes per year, and providing rapid access to the data over regional, continental and transoceanic networks. Computational Grid technology extended to data intensive tasks and worldwide scale could be used to effectively manage such systems. Distributed systems of this size and complexity do not exist yet, although systems of a similar size to those foreseen for the LHC experiments are predicted to come into operation by around 2005.

When the distributed systems are not yet available, as is the case with the ones involved in the future LHC experiments, or their testing implies great costs (and this holds for most of the systems), a simulator becomes a valuable tool for both the system designers and users. The aim of this paper is to describe the simulation program, being developed by the MONARC project (http://monarc.cacr.caltech.edu), as a design and optimization tool for large scale distributed computing systems.

The main goal of the MONARC project is to provide a realistic simulation of large distributed computing systems and to offer a flexible and dynamic environment to evaluate the performance of a range of possible data processing architectures. To achieve this purpose, the simulator provides the mechanisms to describe concurrent network traffic and to evaluate different strategies in data replication or in the job scheduling procedures.

## II. THE SIMULATION FRAMEWORK

A process oriented approach for discrete event simulation is well suited to describe concurrent running programs, network traffic as well as all the stochastic arrival patterns, specific for such type of simulation. Threaded objects or "Active Objects" (having an execution thread, program counter, stack etc.) allow a natural way to map the specific behavior of distributed data processing into the simulation program.

The Java technology was chosen for implementing the simulator as an outcome of the features and advantages that it provides: built-in multithreading support, object-oriented paradigm and portability. This last feature allowed us to test the simulator on a great variety of platforms (mono or multi-processor platforms running Linux, Solaris or Windows).

One of the strengths of MONARC is that it can be easily extended, even by users, and this is made possible by its layered structure. The inner layers contain the core of the simulator (the "simulation engine") and models for the basic components of a distributed system (CPU units, jobs, databases, networks, job schedulers etc.); these are the fixed parts on top of which some particular components (specific for the simulated systems) can be built. These particular components, which form the outer layers, can be different types of jobs, job schedulers with specific scheduling algorithms or database servers that support data replication.

The diagram in Fig. 1 represents the MONARC layers and the way they could interact with a monitoring system.

The simulation framework will be integrated with the MonALISA monitoring system with the purpose of making comparisons between the simulation results and the real systems' behavior.

*A. The Simulation Engine*

The programs running on distributed data processing systems are complex, data dependent and concurrently compete for shared resources. A natural way to simulate their behavior, which was adopted in MONARC, is to use threaded objects ("active objects"), which have an execution thread, program counter, stack and mutual exclusion mechanism. The simulation engine provides a dedicated scheduling mechanism based on semaphores for the active objects.

A base class, `Task`, was created to describe the active objects, and must be inherited by all the entities in the simulation which require a time dependent behavior. Such entities can be the running jobs, the file servers or the database servers. An active object has methods for synchronous and asynchronous communications with other objects, or with the simulation engine, and can be interrupted, suspended and resumed during execution. The behavior of an active object can be a function of messages or data received, its previous state(s), and/or its access to certain shared resources. In this way it is possible to describe highly non-linear processes such as caching and swapping, or the stochastic input pattern for jobs and activities in the system. To allow the communication and the synchronization between active objects we use simulation events. The events implement the interrupt mechanism which describes the sharing of the resources like CPU, memory and I/O between concurrently running tasks. The approach used to simulate the data traffic is also based on an "interrupt" scheme, similar to the one used in the multitasking model. When a new network transfer is initiated, an interrupt is generated and the speed of the affected messages is recomputed. Fig. 2 represents schematically the way this mechanism works.

When a first job (`Task1` from Fig. 2) starts, the time it needs for processing is evaluated (original `TF1`), and the corresponding active object enters into a wait state for this amount of time unless it is interrupted. If a new job (`Task2`) starts on the same hardware, it will cause an interrupt to the first task. Both tasks will share the same CPU power and the time to complete for each of them is re-computed assuming that they share the CPU equally or based on a running priority scheme (new `TF1` and original `TF2`). Then both jobs will enter into a wait state and listen for other interrupts. When the first job (`Task1`) is finished, it creates another interrupt to re-distribute the resources for the remaining jobs. This model assumes that resource sharing is maintained between any discrete events (e.g. new job submission, job completion) that occur during the simulated time interval.

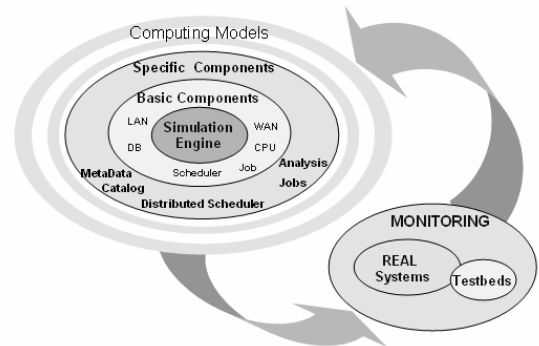

Fig. 1. MONARC layers.

A specific simulation entity, the engine's scheduler, decides when the tasks are allowed to run and is responsible with sending them the appropriate events. It maintains a priority queue with all the future events and, at each simulation step, it extracts from the queue the events with the minimum time stamp and delivers them to the destination tasks. As the number of jobs that must be simulated may be huge, we implemented a dedicated structure that allows active objects recycling in order to improve the simulation efficiency (a pool of active objects).

*B. Basic Components*

The simulation program requires the abstraction of all components from the real systems and their time dependent interactions. This abstracted model has to be equivalent to the original system in the key respects that concern us. The simulation engine is designed to be generic and suitable to describe any distributed system. However, there are certain HEP-specific system components that are especially modeled to make the tool useful for the physics community.

A first set of components was created for describing the physical entities in the simulation. The largest one is the regional center, which contains a farm of working stations (CPU Units), database servers and mass storage units, one or more local area networks. Another set of components describes the behavior of the regional centers and their interaction with the users. Such components are the "Users" or "Activity" Objects which are used to generate data processing jobs based on different scenarios. The job is another base component, simulated with the aid of an active object, and scheduled for execution on a CPU unit by a "Job Scheduler" object.

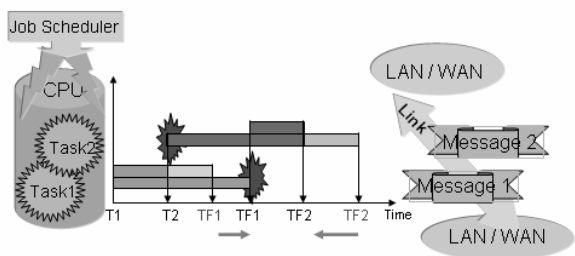

Fig. 2. The interrupt mechanism.

With this structure it is now possible to build a wide range of computing models, from the very centralized (with reconstruction and most analyses at CERN) to the distributed systems, with an almost arbitrary level of complication (CERN and multiple regional centers, each with different hardware configuration and possibly different sets of data replicated).

The main components of a regional center are represented in Fig. 3.

A base component of the MONARC application is the data model. Our data model should provide a realistic mapping of an DBMS, and at the same time allow an efficient way to describe very large database systems with a huge number of objects. For simulating the databases, two main entities used to store data were modeled: the database server and the mass storage center. The database server stores the data on disk drives, while the mass storage center uses tape drives. The users of the distributed system can interact with both those entities, but we also implemented an algorithm that automatically moves the data from a database server to the mass storage server(s) when the first one is out of storage space.

In regard of the network model, the simulation program offers the possibility of simulating data traffic for different protocols on both LAN and WAN. This has to be done for very large amounts of data and without precise knowledge of the network topology (as in the case of long distance connections). It is practically impossible to simulate the networking part at a packet level for such large amounts of data. User defined time dependent functions are used to evaluate the effective bandwidth.

*C. HEP Grid Specific Components*

Beside the basic components described above we have implemented a series of components specific to HEP Grid simulations. This category of components includes a metadata catalog, analysis jobs and a distributed job scheduler.

The metadata catalog is an internal data model component used to store data from the events that are processed in the simulation. It has methods for storing the events with or without creating replicas, for finding the closest database server that holds the data for an event, and for finding the optimum (in terms of network and database costs) database server that holds the data for an event. The metadata catalog is a used in association with jobs that process data in a way similar to how the data are processed in the HEP experiments.

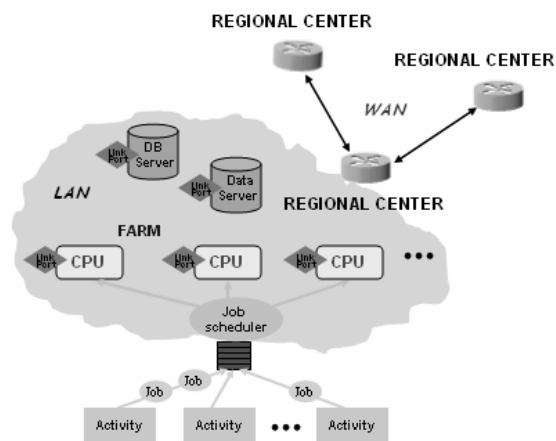

Fig. 3. The regional center components

Another component that is specific to the HEP simulations is the distributed job scheduler. There are two possible solutions for scheduling the jobs on other regional centers than the one they were submitted to: a *centralized algorithm* (the jobs are sent to a global scheduler, which manages the whole system and which will decide where they will run) or a *distributed* one (each local scheduler decides where it is better to send the job). For the time being, we chose the second approach and implemented a distributed job scheduler with a basic algorithm: if the load percentage for each CPU in the local regional center exceeds a certain value, the local scheduler exports the job to another regional center, choosing the one with the minimum average load.

The data analysis jobs are also specific to HEP computing; these jobs use the metadata catalog for obtaining the necessary data, then spend an amount of time with CPU-intensive computations.

III. A SIMULATION STUDY FOR T0/T1 DATA REPLICATION & PRODUCTION ACTIVITIES

*A. INTRODUCTION*

The general concept developed by the two largest experiments, CMS and ATLAS, is a hierarchy of distributed Regional Centers working in close coordination with the main center at CERN. This simulation study follows this concept and describes several major activities;

mainly the data transfer on WAN between the T0 (CERN) and a number of several T1 Regional Centers. The topology describing the connectivity of the Regional Centers is presented in figure 4.

We assume that the three T1 Regional Centers in Europe are connected independently, in a network similar to GEAT. In a simplified model this can be approximated with a "mega-router" in which each T1 regional center is connected through a link. We also consider a transatlantic link connecting T0 with the two T1 regional centers in US and another link connecting the T1 regional centers in Japan. In order to make the files transfer efficient we assume that a transfer Agent runs on all the centers. When it is necessary to send a file to several or all of these centers we have assumed that this is done using the Agent mechanism to provide effective data transfers. In case the same file needs to be transferred to both T1 regional centers in US, the file is transferred only once over the transatlantic line and than copied from T1-US1 to T1-US2 or / and T1-JP.

For the WAN links we assumed the RTT values given in table 1. Those RTT values are used in evaluating the efficiency of using the available bandwidth for "ftp" like transfers.

Using this topology we simulated a number of Activities specific for physics data production, as follows:

1) **RAW Data Replication**.

   From the experiment we assumed a mean rate of recording raw data equal to 200 MB/s. This information is stored in 2GB (normal distributed with 10% sd) data files. These files are replicated in a round robin manner to all 6 T1 regional centers. (The first file is sent to T1-EU1, the second to T1-EU2…)

2) **Production and DST distribution**.
   At T0 all raw data are processed and DST files are generated. The DST files are 10 times smaller in size than the RAW files. We considered again a normal distribution (sd 10%). The DST files created at T0 are sent to all T1 centers. For the T1-US2 and T1-JP the agent transfer system is used to make this operation effective and avoid sending the same file more than once over the same link.

3) **Re-production and new DST distribution**.

   After a certain time the RAW data in each T1 center is re-processed and new DST data is created. Each T1 center will reprocess 1/6 of the RAW data. The DST data generated at each regional center are sent to all the others. Again the agent system is used to effectively transfer data.

4) **Detector Analysis**.

   This activity starts in certain T1 regional centers at given moments of time and collects all RAW data from the other regional centers produced over the last hours. We chose local 9 o'clock as the time this activity starts in the given regional centers and also we chose to gather the RAW data for the last 12 hours. The RAW data is gathered dynamically, meaning from all the regional centers that have the requested data it is chosen the one that maximize the performance of the transfer, based on the network load, proximity and database load.

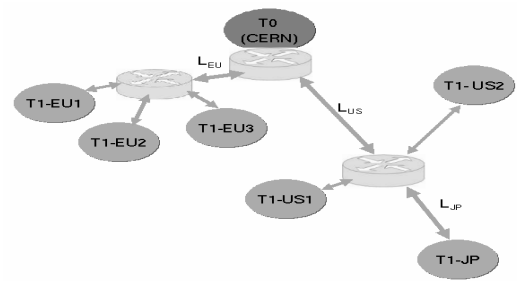

Fig. 4. The network topology considered for the connectivity between the T0 and the T1 Regional Centers

Table 1. RTT values that we used

| Link | RTT (ms) |
|---|---|
| T1-EU1 <-> T0 (CERN) | 20 |
| T1-EU2 <-> T0 (CERN) | 25 |
| T1-EU3 <-> T0 (CERN) | 30 |
| T1-US1 <-> T0 (CERN) | 120 |
| T1-US1 <-> T1-US2 | 60 |
| T1-US1 <-> T1-JP | 240 |

5) **RAW Data Replication**.

   From the experiment we assumed a mean rate of recording raw data equal to 200 MB/s. This information is stored in 2GB (normal distributed with 10% sd) data files. These files are replicated in a round robin manner to all 6 T1 regional centers. (The first file is sent to T1-EU1, the second to T1-EU2…)

6) **Production and DST distribution**.
   At T0 all raw data are processed and DST files are generated. The DST files are 10 times smaller in size than the RAW files. We considered again a normal distribution (sd 10%). The DST files created at T0 are sent to all T1 centers. For the T1-US2 and T1-JP the agent transfer system is used to make this operation effective and avoid sending the same file more than once over the same link.

7) **Re-production and new DST distribution**.

   After a certain time the RAW data in each T1 center is re-processed and new DST data is created. Each T1 center will reprocesses 1/6 of the RAW data. The DST data generated at each regional center are sent to all the other. Again the agent system is used to effectively transfer data.

8) **Detector Analysis**.
This activity starts in certain T1 regional centers at given moments of time and collects all RAW data from the other regional centers produced over the last hours. We choused local 9 o'clock as the time this activity starts in the given regional centers and also we chosen to gather the RAW data for the last 12 hours. The RAW data is gathered dynamically, meaning from all the regional centers that have the requested data it is chosen the one that maximize the performance of the transfer, based on the network load, proximity and database load.

*B. SIMULATION RESULTS*

We simulated the described activities alone and then combined.

We simulated ~1 day of running these activities. In the following figures are some results obtained when running all four activities in parallel. We assumed a mean rate of recording raw data of 200 MB/s. The information is stored in 2GB data files (normal distributed with 10% sd). DST files are produced in the second activities involved at T0 (CERN) from all the RAW data and then are distributed to all the T1 regional centers. The data transfer agent described above is then used. After a certain period of time each T1 center will start to re-process the raw data stored locally and to generate a new set of DST. Each T1 has ~1/6 from the entire raw data and will generate new DST which should be replicated to all the other regional centers. As before, in this case we also assume that transfer agents are running on all the centers involved (T0, T1-US1) for an effective replication. Finally, the Detector Analysis activity runs on T1-JP regional center and starts at 9 o'clock local time. Then it will gather the RAW data produced in the last 12 hours from the others centers using a get-optimum-performance algorithm as mentioned above.

Using this configuration we did a series of tests in which we have varied the available bandwidth between T0 (CERN) and T1-US1. In the following figures are the obtained results.

In the figure 5 is the representation of how varies the time with which the DST files are served in different T1 centers for the test cases in which the available bandwidth between T0 (CERN) and T1-US1 varies between 3Gbps and 10Gbps. As seen the DST files transfer time tends to decrease proportionally with the amount of bandwidth available between T0 (CERN) and T1-US1 centers. The series "all Series" represents the average value of the DST files transfer time considering all the T1 tiers in the simulation.

In the figure 6 is the representation of the way the RAW file transfer time varies in different T1 centers in the tests in which we have varied the amount of available bandwidth between T0 (CERN) and T1-US1.

As seen the RAW files transfer time tends to decrease proportionally with the amount of bandwidth available between T0 (CERN) and T1-US1 centers. The series "all Series" represents the average value of the RAW files transfer time considering all the T1 tiers in the simulation.

In the figure 7 is the representation of the variation of time needed to complete the Detector Analysis activity in the tests done for different values for the amount of available bandwidth between T0 (CERN) and T1-US1. As said above this activity gathers the RAW data from the last 12 hours, but as seen here when using a 3Gbps link it takes almost 24 hours to finish, while when using a 10Gbps link between T0 (CERN) and T1-US1 it takes around 15 hours to finish.

We will present as follows a comparison for production and DST distribution, done with and without the Data Transfer Agent.

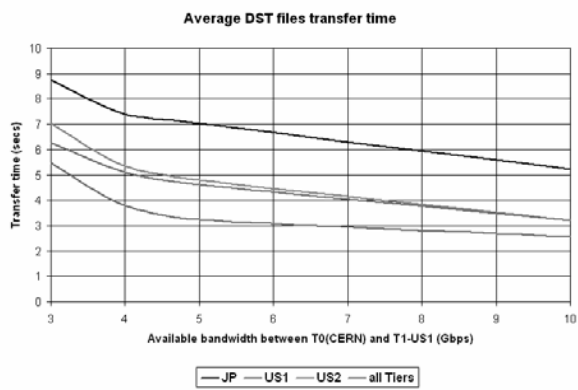

Fig. 5. The DST files transfer time in different T1 centers with different values for the available bandwidth between T0 (CERN) and T1-US1

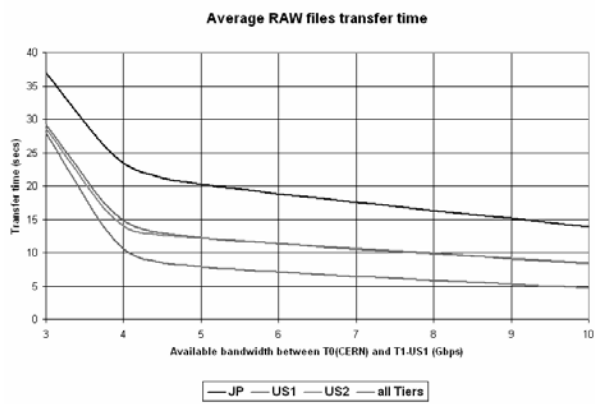

Fig. 6. The RAW files transfer time in different T1 centers with different values for the available bandwidth between T0 (CERN) and T1-US1

In the Production and DST distribution activity test at T0 (CERN) regional center are produced DST files from the recorded RAW data, which are then distributed to all the T1 regional centers. The Data Transfer Agent is used on the T1-US1 regional center and will forward the DST data received in that center from T0 (CERN) further to T1-US2 and T1-JP regional centers. This means that at T0 (CERN) when using the Data Transfer Data the DST files will be sent only to T1-EU regional centers and to T1-US1, while the agent will handle the further transfer of those files from T1-US1 to the rest of the regional centers.

For the average used bandwidth on the major links the obtained results are shown in figures 8 and 9. As seen the average bandwidth used on the CERN link is greater when we do not use the Data Transfer Agent since more data get transferred from that regional center.

## IV. SIMULATION STUDY: PROOF CLUSTER

*A. Introduction*

We are currently writing a series of simulation examples with the purpose of modeling the behavior of Proof clusters. Proof is a facility for distributed data processing under the Root framework, developed at CERN (for more details see http://root.cern.ch).

A Proof configuration consists of several clusters; the computers from a cluster run master and slave processes, as shown in Fig. 10.

The typical scenario for data processing with Proof contains the following phases:

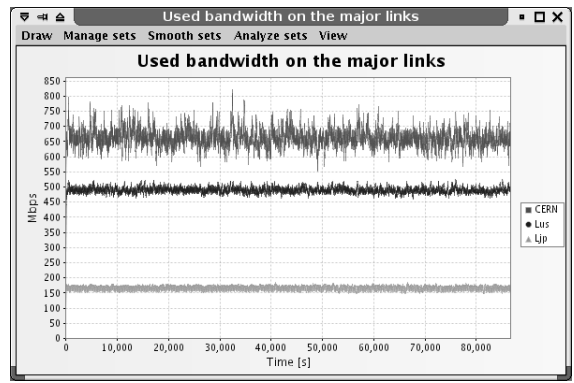

Fig. 8. The used bandwidth on the major links output obtained for the test done using the Data Transfer Agent

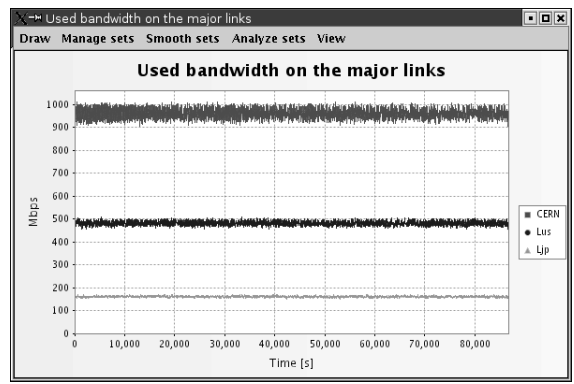

Fig. 9. The used bandwidth on the major links output obtained for the test done without using the Data Transfer Agent

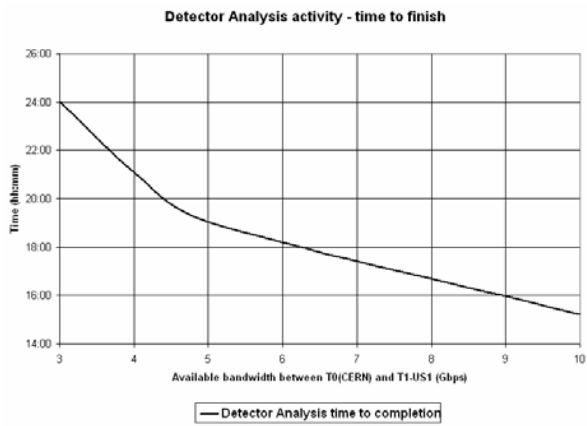

Fig. 7. Time needed for the Detector Analysis activity to finish for the tests done centers with different values for the available bandwidth between T0 (CERN) and T1-US1

- a client sends a request to a master, specifying a dataset to be processed
- the master identifies the files that contain the needed data and determines their location;
- the master identifies the files that contain the needed data and determines their location;
- each slave enters a loop in which it asks the master for a work packet (which specifies a number of events to be processed), it executes the task and sends the result back to the master
- the master assigns work to the slaves taking into account the location of the files and the relative performance of the slaves

There are three possibilities for the slaves to obtain the data they are assigned: from the local disk, from a server or from other slave stations, with the aid of the *rootd* server (*rootd* is a daemon that allows remote access to Root database files). All these three possibilities can be modeled by the generic diagram from Fig. 10.

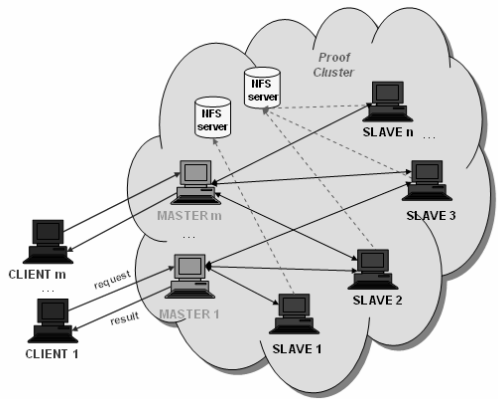

Fig. 10. Proof cluster architecture.

*B. Example Description*

The simulated scenario is based on the general description presented above. The working cluster contains $n$ master stations, $m$ slave stations and $s$ data servers (we tested with $n=20$ and $m=500$, and with different values for $s$). Each master receives a data processing request from a client; we assumed that the client needs to process a set of files containing analysis data for a certain number of events. We also tested some cases in which the clients repeatedly send requests to the masters, with pause intervals between requests.

When asking the master for work, a slave is assigned one file which is assumed to be available on the slave's local disk with a certain probability; if not available, the file is taken from a data server. It was assumed that the master takes some time to handle a work request from a slave and to process the partial results returned by a slave; if there are several slaves that send work requests or partial results at the same time, their messages will be processed by the master sequentially.

The behavior of the system was studied by varying several parameters such as:

- the number of slave processes created by each master (on a slave station there can be more than one slave process; in our test cases, the minimum number of processes on a slave station was 1, corresponding to 25 slave processes created by each master - as we have 20 masters and 500 slave nodes)
- the probability of having the data on the local disk at the slave nodes
- the LAN bandwidth

*C. Simulation Results*

There were several test cases in which we obtained a substantial throughput improvement in the simulated system: when the LAN bandwidth was increased (from 100Mbps to 1Gbps and even 500Mbps), when the processing time at the data server was reduced and when some additional data servers containing replicated data were introduced.

The advantage of having several slave processes on a machine is that while some of them are waiting for data from the network, the others can do CPU-intensive operations. However, there is also a disadvantage of having more slave processes: they can create network bottlenecks and waiting queues at the data server and at the master (when requesting work packets and sending results). The graph from Fig. 11 represents the total processing time of a constant number of jobs in three situations: when each master creates 25, 50 and 100 slave processes. In this test case, when the data server is single threaded, the optimum number of slaves is the smallest one (25).

As mentioned above, we also simulated the situation in which the clients send several requests to the masters, with breaks between requests. An average request would take about 2-3h to be processed on a single CPU, but only takes a few minutes to be processed in the cluster. The breaks between requests also have the average value of 5min.

In this case, having more slave processes on a station leads to a better throughput, beacuse the station has a greater probability of being active even if not all the masters are processing a request at that moment. We simulated test cases with 25, 50 and 100 slave processes created by each master and computed the average CPU usage for different local data probabilities. These average values, for a 500Mbps network bandwidth, are shown in Fig. 12.

Fig. 13 is part of the simulation output and presents the CPU utilization in the cluster for a test case with 50 slave processes.

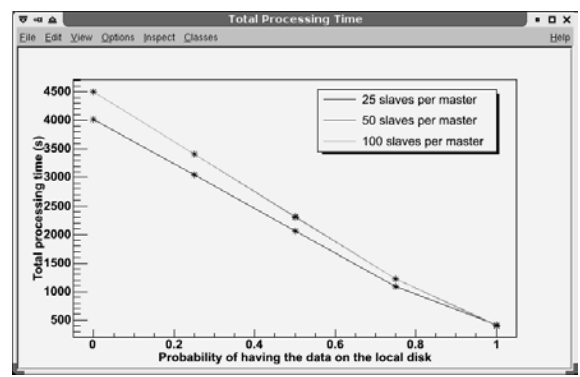

Fig. 11. Total processing time with different numbers of slave processes.

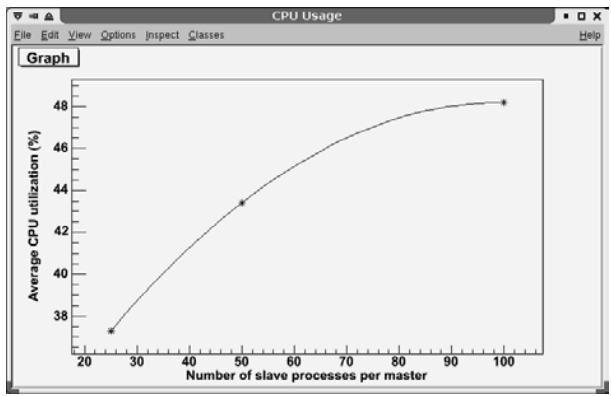

Fig. 12. Average CPU usage in three test cases: 25, 50 and 100 slave processes per master.

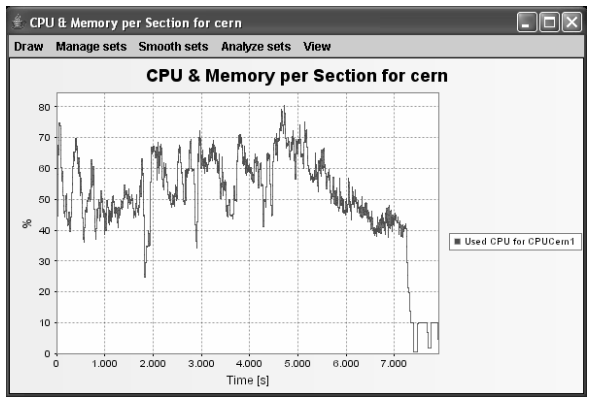

Fig. 13. CPU and bandwidth utilization on the slave nodes (test case with 50 slave processes per master, 50% local data probability).

## V. SUMMARY

A CPU and code-efficient simulation approach to the problem of simulation of distributed computing systems has been developed and tested within the MONARC Collaboration. It provides a transparent way to map the distributed data processing, data transport and analysis tasks onto the simulation frame, and can describe dynamically even very complex computing models. The Java(TM) programming environment, used extensively to build the MONARC simulation tool, is very well suited for developing a flexible and distributed process oriented simulation, equipped with adequate graphical and statistical tools. This simulation program is still under development to include more sophisticated methods to evaluate different strategies to optimize the utilization of resources in very large scale distributed computing systems.

### ACKNOWLEDGMENTS

The authors wish to thank Harvey B. Newman, Iosif C. Legrand from Caltech and Nicolae Tapus and Valentin Cristea from UPB, as well as all MonALISA team.